\documentclass[12pt,preprint]{aastex}

\begin{document}
\def\Msun{M_{\odot}}
\def\Lsun{L_{\odot}}
\def\noi{\noindent}

\setcounter{page}{0}

\begin{center}
{\Large {\bf Dwarf Galaxies in 2010: Revealing Galaxy Formation's Threshold and Testing the Nature of Dark Matter}}

James S. Bullock$^\dagger$ and Manoj Kaplinghat \\
{\em Physics \& Astronomy Department, University of California, Irvine;} $^\dagger$bullock@uci.edu \\
\bigskip
Andrew Fruchter \\
{\em Space Telescope Science Institute} \\
\bigskip
Marla Geha \\
{\em Astronomy Department, Yale University} \\
\bigskip
Joshua D. Simon \\
{\em Observatories of the Carnegie Institution of Washington}   \\
\bigskip
Louis E. Strigari \\
{\em Physics Department, Stanford University} \\
\bigskip
Beth Willman \\
{\em Astronomy Department, Haverford College} \\

\end{center}

\begin{center}
\includegraphics[width=0.9\textwidth]{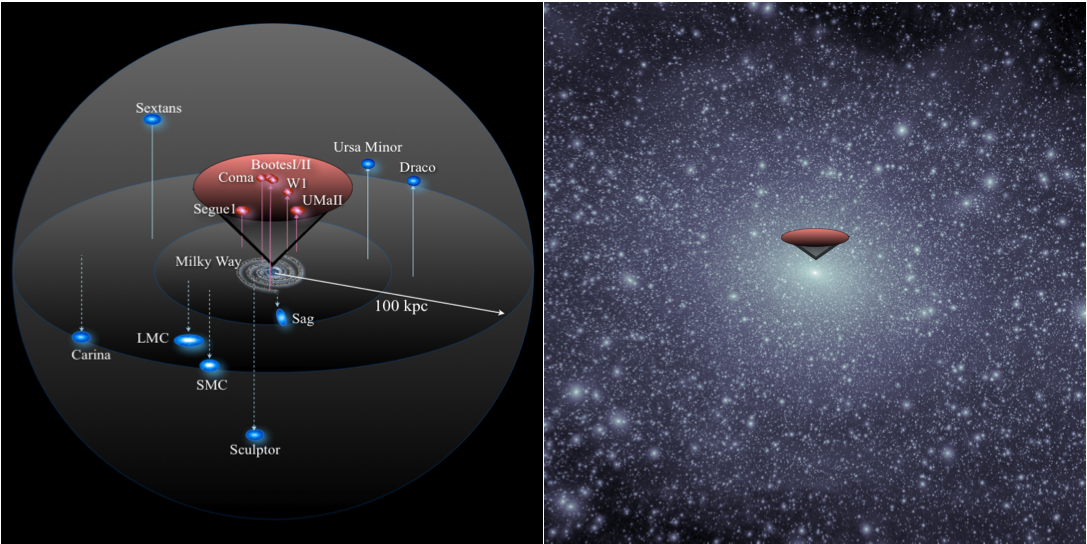}
\end{center}
{{ ({\em Left}) Known satellite galaxies of the Milky Way within $100$ kpc.
({\em Right}) LCDM simulation of a Milky Way size halo  (Diemand et al. 2008; spanning $800$ kpc).\nocite{diem:08}
The red-shaded cones represent the volume that is currently complete for
discovery and characterization of ultra-faint dwarf galaxies.}}

\bigskip
\noi \underline{Panels}: GAN and CFP

\newpage

\begin{center}
{\bf Abstract}
\end{center}
Over the past five years, searches in Sloan Digital Sky Survey data have more than doubled the number of known dwarf satellite galaxies of the Milky Way, and have revealed a population of ultra-faint galaxies with luminosities smaller than typical globular clusters, $L \sim $ 1000 L$_\odot$.  These systems are the faintest, most dark matter dominated, and most metal poor galaxies in the universe.  
Completeness corrections suggest that we 
are poised on the edge of a vast discovery space in galaxy phenomenology, with hundreds more of these extreme galaxies to be discovered as future instruments hunt for the low-luminosity threshold of galaxy formation.
Dark matter dominated dwarfs of this kind probe the small-scale power-spectrum, provide the most stringent limits on the phase-space packing of dark matter,
 and offer a particularly useful target for dark matter indirect detection experiments.    Full use of dwarfs as dark matter laboratories will
require synergy between  deep, large-area  photometric searches; spectroscopic  and astrometric follow-up with 
next-generation optical telescopes; and subsequent observations with gamma-ray telescopes for dark 
matter indirect detection. 

\begin{center}
{\bf 1. Introduction and Outline}
\end{center}

\noi  Since 2004, twenty-five new dwarf galaxy companions of the Milky Way and M31
have been discovered, with most of them less luminous than any galaxy previously
known  \cite[e.g.,][]{willman:05,belo:07,m:08}.  The most extreme ultra-faint dwarfs have luminosities smaller than the average globular cluster (L$_{V} \simeq 10^3 - 10^4$ L$_\odot$)  and can only be detected as slight overdensities of resolved stars in deep imaging surveys. Follow-up spectroscopy reveals that these, 
the faintest galaxies known, are also the most dark matter
dominated \citep{martin07,sg07,strig:08a} and most metal poor \citep{kirby08,geha08} stellar systems yet observed.

Perhaps the most exciting aspect of these recent discoveries is that they hint at a much
larger population. As discussed in \S 2 and illustrated in Figure 1, detection is complete only to $\sim 50$ kpc for  the faintest
dwarfs \citep{kopo:08,walsh09}.  Straightforward 
luminosity bias corrections suggest as many as $\sim 500$ ultra-faint dwarf galaxies within
the virial radius of the Milky Way (Tollerud et al. 2008). \nocite{toll:08}
The implication is that we are poised at the edge of a vast new discovery space in galaxy phenomenology.
Deep searches for these galaxies will explore galaxy formation's threshold, 
and follow-up studies will allow us to test ideas about galaxy formation suppression and chemical 
enrichment in the smallest dark matter halos at the earliest times
\citep{bkw00,kravtso_etal04,stri:07a,Li08,maccio09,kopo09,busha09}.

The discovery of these nearby dark matter dominated galaxies provides a great 
opportunity to understand the microscopic nature of dark matter and test the
Cold Dark Matter (CDM) paradigm.
As discussed in \S 3, 
Local Group dwarf spheroidal galaxies (dSphs) 
are ideally suited for this purpose: 1) they are close enough
to be studied using kinematics of {\em individual stars}; 2) their high dark matter fractions ($M/L > 100$)
render them relatively
free of the usual baryonic uncertainties that make dark matter profile
determinations difficult; and 3) they inhabit the smallest dark matter halos known to host stars
 and therefore provide the best direct laboratories for testing dark matter clustering
properties on small scales -- scales where CDM theory faces its most serious challenges.

Dwarf galaxies are also excellent laboratories for indirect detection of dark
matter.
As discussed in \S 4,  if the dark matter is of the   WIMP   variety,  then dark
  matter
annihilations can  produce potentially-detectable gamma-ray
photons    \citep[see   e.g., ][and  references   therein]{strig:08b}.
Indirect dark matter detection of this kind is one of the major goals
of current and proposed gamma ray telescopes.
  Unlike the Galactic Center, 
dSphs are devoid of gas and ongoing star formation, and therefore contamination from astrophysical gamma-ray sources like pulsars will be minimal.  If dark matter self-annihilates with weak-scale cross sections (as many theories predict) then dwarf satellites will complement the Galactic Center to provide the most robust means for dark matter detection.
In order to turn such a detection (or upper limits on the flux) into
constraints on particle physics models, we need accurate dynamical mass models
for the dwarfs. This fact points to
an important
synergy between deep, wide field photometric searches for new dwarf galaxies, 
optical kinematic follow-up for dark matter mass models, and subsequent observations with
gamma-ray telescopes for dark matter indirect detection studies.

\begin{figure}[bt]
\begin{center}
\includegraphics[width=0.95\textwidth]{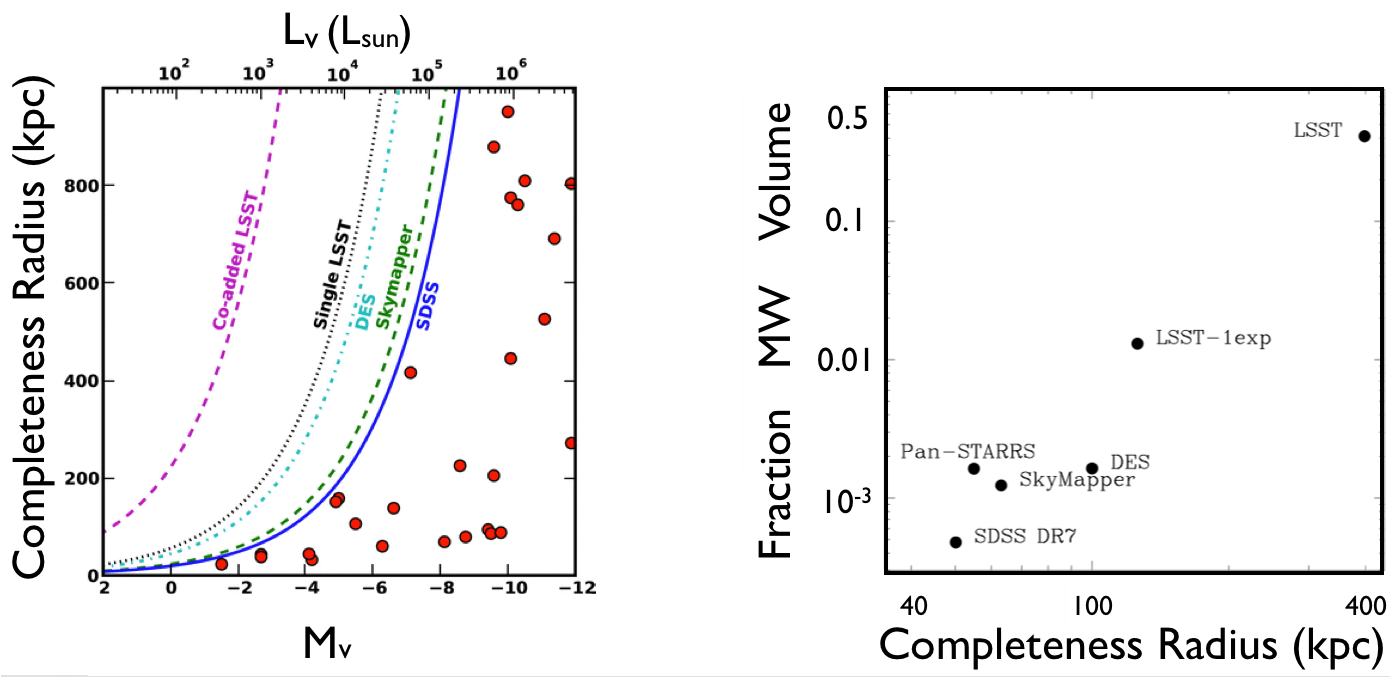}
\end{center}
\vskip-0.25cm
\caption{\small{({\it Left}) 
The radius out to which detection is complete for faint dwarf galaxies in various surveys (Tollerud et al. 2008).
  Known dwarfs are indicated as red circles, which trace the completeness threshold for SDSS.    ({\it Right}) The fraction of the Milky Way halo volume (out to 400 kpc) within  which a given survey can detect $L \sim 10^3 L_\odot$ dwarfs as  a function of completeness radius.}}\label{fig_surveys}
\end{figure}

\begin{center}
{\bf 2. Hunting for the Low-Luminosity Threshold of Galaxy Formation}
\end{center}

\noi  The recent discoveries of dark matter dominated galaxies with luminosities as low as $\sim 10^3 \Lsun$ 
raise a number of fundamental questions.   Is there a low-luminosity threshold for galaxy formation?
If so, what physics sets this scale?  Is there a dark matter halo mass
scale below which galaxy formation is completely suppressed?
What are the chemical enrichment histories for galaxies
that are themselves less luminous than a star cluster?  While it is difficult to answer
these questions with current facilities, the future is promising.

 By carefully filtering star catalogs, based on color and magnitude, it is
possible to find ultra-faint galaxies as overdensities of resolved
stars.  This technique 
requires homogeneous photometric coverage over a large portion of the
sky.  The Sloan Digital Sky Survey (SDSS) is the first survey with
sufficient depth and sky coverage to perform this test and all of 
the Milky Way ultra-faint galaxies have so far
been discovered in this dataset.
The left panel of Figure 1 illustrates the radius out to which dwarf
galaxies of a given luminosity can be discovered with several proposed surveys.  Known dwarfs are red circles.  
The right panel shows the fraction of the Milky Way halo's volume in which
ultra-faint dwarf galaxies may be detected, as a function of the
associated radius of completeness.   We see that {\bf LSST, covering 
nearly half the sky, will be able to detect ultra-faint dwarf galaxies out beyond the edge
of the Milky Way's virial radius.}

Reasonable predictions for the total number of luminous Milky Way
satellites range between 60 -- 600, depending on the physics of
low-mass galaxy formation (and associated radial distribution of dwarfs). 
The new galaxies that should be discovered by future surveys will greatly
influence our interpretation of the so-called `Missing Satellites Problem' in
CDM \citep{klypin99}. The census of dwarf galaxies in the Milky
Way volume will also provide a robust upper limit on the free-streaming
 scale of the dark matter particle --   warm dark matter models are ruled out if they
  predict fewer subhalos than observed satellite galaxies
  
 \begin{center}
{\bf 3. Taking Dark Matter's Temperature}
\end{center}

\noi Photometric surveys are necessary to discover new faint dwarf galaxies. Once
found, \emph{we can learn about the nature of these objects and the dark
matter within them only through spectroscopic followup}.  With medium-resolution
spectroscopy of stars in dwarf galaxies, it is possible to measure 
velocities of individual stars (good to a few km~s$^{-1}$) and hence the masses of dwarf galaxies.
Over the past 20 years, these measurements have revealed that dwarf galaxies ---
including the newly discovered ultra-faint dwarfs --- are highly dark
matter-dominated systems, with central mass-to-light ratios approaching 1000 in solar
units \citep[e.g.,][]{aaronson83,mateo93,sg07,geha08}.

One of the most striking recent findings about Milky Way satellite
galaxies was the discovery that \emph{all nineteen} dSphs, covering
more than four orders of magnitude in luminosity, inhabit dark matter
halos with the same mass ($\sim10^{7}$~M$_{\odot}$) within their
central 300~pc \citep[][]{strig:08a}.  This result is shown in Figure 2.
It suggests either a
lower mass limit for galaxy formation, or conceivably a cutoff in the
mass function of dark matter halos, which would map to the properties
of dark matter particles.  Improved mass measurements (utilizing much
larger samples of stellar velocities) and detailed studies of the
lowest luminosity objects are urgently needed to determine more
accurately the mass-luminosity relationship and verify the
non-existence of galaxies with masses below the apparent threshold.
Unfortunately, confirming whether stellar overdensities  fainter
than $M_{V} \sim -4$ are truly dark matter dominated dwarfs
is already a significant challenge because of
the difficulty of identifying a small number of kinematically
associated stars in the face of a large Milky Way foreground
population covering a range of velocities. {\bf  Future instruments such
as a wide-field spectrograph on an extremely large ground-based telescope
(providing a factor of $\sim10$ improvement in collecting area
compared to Keck) or WFMOS on an 8~m telescope 
would tremendously enhance our ability to study
dark matter and galaxy formation with ultra-faint dwarfs.}

\begin{figure}[bt]
\begin{center}
\includegraphics[width=0.95\textwidth]{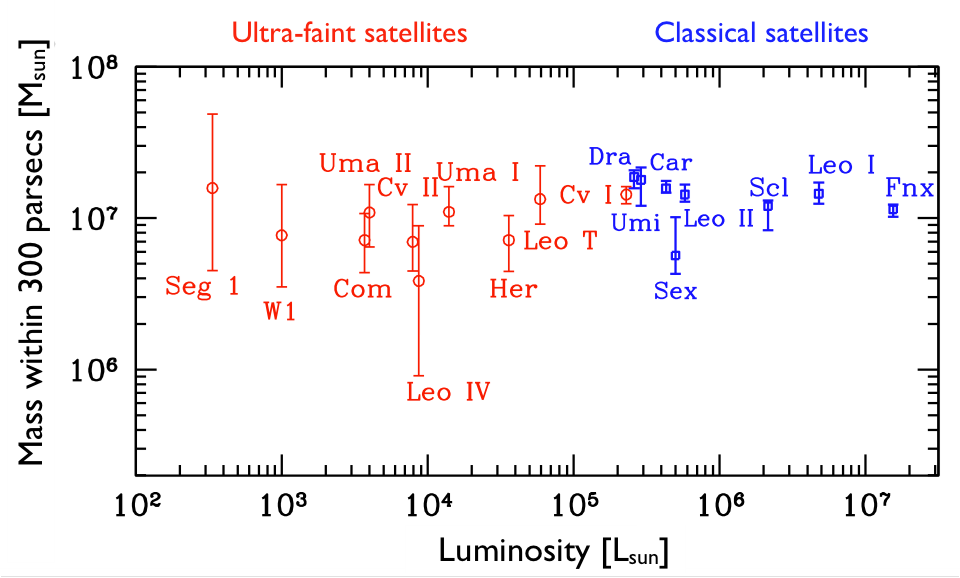}
\end{center}
\vskip-0.25cm
\caption{\small{The integrated mass within the inner 300pc of Milky Way satellite galaxies as a function 
of their luminosities (Strigari et al. 2008).  Satellites discovered since 2004 are shown in red.}}\label{fig_surveys}
\end{figure}

An important area where dwarf galaxy kinematic studies will 
be useful is in placing limits on (or measuring the existence of) a 
phase-space limited core in their dark matter halos.  This will provide an important
constraint on the nature of dark matter.  The ability for dark matter to pack
in phase space is limited by its intrinsic properties such as mass and formation
mechanism. CDM particles have negligible
velocity dispersion and very large central phase-space density,
resulting in cuspy density profiles over observable scales
\citep{NFW}.
Warm Dark Matter (WDM) halos, in contrast, have smaller central
phase-space densities, so that density profiles saturate to
form constant central cores.   Owing to their small masses, dSphs 
have the highest average
phase space densities of any galaxy type, 
and this implies that for a given DM model,
phase-space limited cores will occupy
a larger fraction of the virial radii.  This means that dSphs are the
most promising galaxy candidates for manifesting phase-space cores.
Unfortunately, kinematic measurements of the {\em stars} in galaxies
provide only a lower limits on the coarse-grained
phase space density of the {\em dark matter particles} in those galaxies.
The only way to detect a phase-space core within a dark matter halo is to
detect the presence of a constant-density core in its density profile.

\begin{figure}[t!]
\vskip 5mm
\begin{center}
\includegraphics[width=0.95\textwidth]{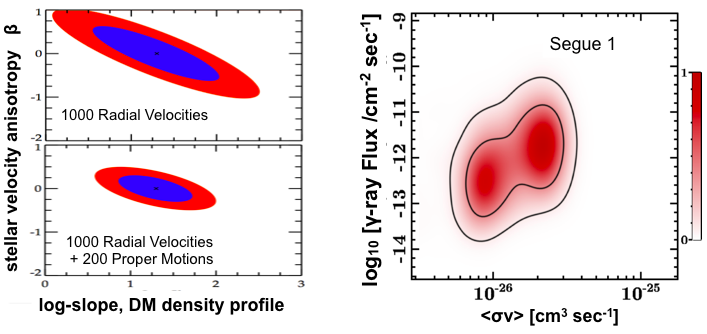}
\end{center}
\begin{tiny}
\caption{ ({\em Left})  Errors in the 
measured stellar velocity anisotropy vs. dark halo density profile slope for a simulated dSph. 
Upper panel shows the implied errors for 1000 radial velocities at $5$ km s$^{-1}$ accuracy and lower panel includes 200 proper motions at 15 $\mu$as yr$^{-1}$.  We see an experiment like SIM Lite
would be able to discriminate between a cusp and a core.
({\em Right}) Likelihood of predicted flux of $E > 1$GeV gamma rays from the dwarf galaxy Segue 1, based on a
joint marginalization over
 Constrained Minimal Supersymmetric Standard Model parameter space and the
current kinematic data for this dwarf \citep{Martinez09}.}
\end{tiny}
\vskip 3mm
\label{BETA_SLOPE}
\end{figure}

Current observations are unable to measure the density profile slopes
within dSph galaxies because of a strong degeneracy between the inner slope
of the DM density profile and the velocity anisotropy,
$\beta$, of the stellar orbits.
Radial velocities alone cannot break this
degeneracy (Fig.~3),
even if the present samples of radial velocities are
increased to several thousand stars \citep{Strig:07b}.
Combining {\em proper motions} with the present
samples of radial velocities will provide orthogonal constraints
and is the only robust means of breaking the anisotropy-inner slope degeneracy.
The most promising dSphs for this experiment will be nearby and high surface brightness systems such as Fornax and Sculptor, for which the upper giant branches require proper 
motions of stars with $V \simeq 19$.
The needed measurements will include
proper motions for $\sim 100$ stars per galaxy with accuracies better
than 10 km s$^{-1}$ ($< 40 \mu$as yr$^{-1}$ at 60 kpc).
Figure 3 shows
that with about 1000 radial velocities and 200 transverse velocities
of this accuracy,
it will be possible to reduce the error on the log-slope of the dark
matter density profile to about 0.2 \citep[][]{Strig:07b}. This will enable
us to rule out a large class of warm dark matter models or invalidate the cold
dark matter paradigm. 
{\bf The required observations are well-matched to the projected performance of SIM Lite}
but well-beyond the capabilities of Gaia.   This kind of measurement may also be
possible with giant ground-based or space telescopes (like TMT, GMT, and eventually ATLAS-T)
with repeated observations over a few years. Astrometry at the 
100~$\mu$as level has already been demonstrated with a laser guide 
star adaptive optics system on the Palomar 5~m telescope, and extrapolating 
those results to a 25~m or 30~m aperture indicates an expected limit of 10~$\mu$as \citep{cameron09}. 
The increased number of radial velocities made possible with 30 m telescopes will also reduce
the associated errors on the log-slope compared to those presented in Fig. 3.

\begin{center}
{\bf 4. Indirect Detection of Dark Matter}
\end{center}

\noindent Detecting dark matter through the products of its decay or self-annihilation in an astrophysical system is an exciting prospect.
Moreover it is possibly the {\em only way} we can infer or confirm the nature of {\em all} of the dark matter in the universe. Dark matter models from theories with new physics at the weak scale (the scale the LHC is designed to probe) generically predict high-energy annihilation products such as gamma-rays. The flux of gamma-rays depends both on the dark matter particle properties as well the spatial distribution of dark matter in the astrophysical source of photons. The closest and densest
dwarf galaxies are expected to be the brightest 
sources \citep{strig:08b,kuhlen08} after the Galactic Center, so identifying the
most promising targets and predicting the flux ratios between different
objects strongly motivates extensive stellar velocity measurements of
the lowest luminosity systems.  

Kinematics in dark matter dominated dwarf satellites provide robust
constraints on the spatial distribution of dark matter in them.
A complicating factor here is dark matter substructure in dwarf galaxies.
Substructure does not contribute significantly to the dynamical mass but because of its
elevated density ``boosts'' the annihilation signal. Recent work has shown
that using a combination of numerical simulations \citep{Springel08} and
analytic modeling \citep{Martinez09}, it is possible to account
for the substructure contribution in flux calculations.
Figure 3 shows such a calculation for the 
 predicted gamma-ray flux above 1 GeV (appropriate for the Fermi satellite observatory) versus annihilation cross-section for Segue 1, a new Milky Way satellite at 23 kpc. The particle model assumed here is the Constrained Minimal Supersymmetric
Standard Model~\citep[CMSSM, ][]{Austri06} including 
WMAP and current accelerator constraints. Most of the
likely flux values for the CMSSM are, however, below the sensitivity
limit of the Fermi observatory.

Imaging Air-Cerenkov Telescopes (IACTs) such as HESS, MAGIC and VERITAS have angular resolutions of $\sim 0.1^\circ$ at high energies, and allow pointed observations of dwarf satellites.  Present limits from ACT observations of dwarf satellites are unable to probe down to the WMAP-favored CMSSM parameter space \citep{Aharonian2008,Aliu2008,Wood2008}.  
However, a 1 km$^2$ IACT instrument~\citep{Buckley:2008zc} with better background rejection techniques 
should get down to these sensitivities.  A facility of this type holds real promise for detecting dark matter through non-gravitational means.

\bibliographystyle{apj}

\bibliography{Astro2010_dwarf}

\begin{thebibliography}{34}
\expandafter\ifx\csname natexlab\endcsname\relax\def\natexlab#1{#1}\fi

\bibitem[{{Aaronson}(1983)}]{aaronson83}
{Aaronson}, M. 1983, \apjl, 266, L11

\bibitem[{Aharonian {et~al.}(2008)}]{Aharonian2008}
Aharonian, F., {et~al.} 2008, Astropart. Phys., 29, 55

\bibitem[{{Aliu, E. et al.}(2008)}]{Aliu2008}
{Aliu, E. et al.} 2008, arXiv e-prints 0810.3561

\bibitem[{{Belokurov} {et~al.}(2007){Belokurov}, {Zucker}, {Evans}, {Kleyna},
  {Koposov}, {Hodgkin}, {Irwin}, {Gilmore}, {Wilkinson}, \& {et~al.}}]{belo:07}
{Belokurov}, V., {Zucker}, D.~B., {Evans}, N.~W., {Kleyna}, J.~T., {Koposov},
  S., {Hodgkin}, S.~T., {Irwin}, M.~J., {Gilmore}, G., {Wilkinson}, M.~I., \&
  {et~al.} 2007, \apj, 654, 897

\bibitem[{Buckley {et~al.}(2008)}]{Buckley:2008zc}
Buckley, J., {et~al.} 2008, arXiv e-prints 0812.0795

\bibitem[{{Bullock} {et~al.}(2000){Bullock}, {Kravtsov}, \& {Weinberg}}]{bkw00}
{Bullock}, J.~S., {Kravtsov}, A.~V., \& {Weinberg}, D.~H. 2000, \apj, 539, 517

\bibitem[{{Busha} {et~al.}(2009){Busha}, {Alvarez}, {Wechsler}, {Abel}, \&
  {Strigari}}]{busha09}
{Busha}, M.~T., {Alvarez}, M.~A., {Wechsler}, R.~H., {Abel}, T., \& {Strigari},
  L.~E. 2009, ArXiv e-prints 0901.3553

\bibitem[{{Cameron} {et~al.}(2009){Cameron}, {Britton}, \&
  {Kulkarni}}]{cameron09}
{Cameron}, P.~B., {Britton}, M.~C., \& {Kulkarni}, S.~R. 2009, \aj, 137, 83

\bibitem[{{Diemand} {et~al.}(2008){Diemand}, {Kuhlen}, {Madau}, {Zemp},
  {Moore}, {Potter}, \& {Stadel}}]{diem:08}
{Diemand}, J., {Kuhlen}, M., {Madau}, P., {Zemp}, M., {Moore}, B., {Potter},
  D., \& {Stadel}, J. 2008, \nat, 454, 735

\bibitem[{{Geha} {et~al.}(2008){Geha}, {Willman}, {Simon}, {Strigari}, {Kirby},
  {Law}, \& {Strader}}]{geha08}
{Geha}, M., {Willman}, B., {Simon}, J.~D., {Strigari}, L.~E., {Kirby}, E.~N.,
  {Law}, D.~R., \& {Strader}, J. 2008, ArXiv e-prints 0809.2781

\bibitem[{{Kirby} {et~al.}(2008){Kirby}, {Simon}, {Geha}, {Guhathakurta}, \&
  {Frebel}}]{kirby08}
{Kirby}, E.~N., {Simon}, J.~D., {Geha}, M., {Guhathakurta}, P., \& {Frebel}, A.
  2008, \apjl, 685, L43

\bibitem[{{Klypin} {et~al.}(1999){Klypin}, {Kravtsov}, {Valenzuela}, \&
  {Prada}}]{klypin99}
{Klypin}, A., {Kravtsov}, A.~V., {Valenzuela}, O., \& {Prada}, F. 1999, \apj,
  522, 82

\bibitem[{{Koposov} {et~al.}(2008){Koposov}, {Belokurov}, {Evans}, {Hewett},
  {Irwin}, {Gilmore}, {Zucker}, {Rix}, {Fellhauer}, {Bell}, \&
  {Glushkova}}]{kopo:08}
{Koposov}, S., {Belokurov}, V., {Evans}, N.~W., {Hewett}, P.~C., {Irwin},
  M.~J., {Gilmore}, G., {Zucker}, D.~B., {Rix}, H.-W., {Fellhauer}, M., {Bell},
  E.~F., \& {Glushkova}, E.~V. 2008, \apj, 686, 279

\bibitem[{{Koposov} {et~al.}(2009){Koposov}, {Yoo}, {Rix}, {Weinberg},
  {Macci{\`o}}, \& {Miralda-Escud{\'e}}}]{kopo09}
{Koposov}, S.~E., {Yoo}, J., {Rix}, H.-W., {Weinberg}, D.~H., {Macci{\`o}},
  A.~V., \& {Miralda-Escud{\'e}}, J. 2009, ArXiv e-prints 0901.2116

\bibitem[{{Kravtsov} {et~al.}(2004){Kravtsov}, {Gnedin}, \&
  {Klypin}}]{kravtso_etal04}
{Kravtsov}, A.~V., {Gnedin}, O.~Y., \& {Klypin}, A.~A. 2004, \apj, 609, 482

\bibitem[{{Kuhlen} {et~al.}(2008){Kuhlen}, {Diemand}, \& {Madau}}]{kuhlen08}
{Kuhlen}, M., {Diemand}, J., \& {Madau}, P. 2008, \apj, 686, 262

\bibitem[{{Li} {et~al.}(2008){Li}, {Helmi}, {De Lucia}, \& {Stoehr}}]{Li08}
{Li}, Y.-S., {Helmi}, A., {De Lucia}, G., \& {Stoehr}, F. 2008, ArXiv e-prints
  0810.1297

\bibitem[{{Macci{\`o}} {et~al.}(2009){Macci{\`o}}, {Kang}, \&
  {Moore}}]{maccio09}
{Macci{\`o}}, A.~V., {Kang}, X., \& {Moore}, B. 2009, \apjl, 692, L109

\bibitem[{{Martin} {et~al.}(2007){Martin}, {Ibata}, {Chapman}, {Irwin}, \&
  {Lewis}}]{martin07}
{Martin}, N.~F., {Ibata}, R.~A., {Chapman}, S.~C., {Irwin}, M., \& {Lewis},
  G.~F. 2007, \mnras, 380, 281

\bibitem[{{Martinez} {et~al.}(2009){Martinez}, {Bullock}, {Kaplinghat},
  {Strigari}, \& {Trotta}}]{Martinez09}
{Martinez}, G., {Bullock}, J.~S., {Kaplinghat}, M., {Strigari}, L.~E., \&
  {Trotta}, R. 2009, in Preparation

\bibitem[{{Mateo} {et~al.}(1993){Mateo}, {Olszewski}, {Pryor}, {Welch}, \&
  {Fischer}}]{mateo93}
{Mateo}, M., {Olszewski}, E.~W., {Pryor}, C., {Welch}, D.~L., \& {Fischer}, P.
  1993, \aj, 105, 510

\bibitem[{{McConnachie} {et~al.}(2008){McConnachie}, {Huxor}, {Martin},
  {Irwin}, {Chapman}, {Fahlman}, {Ferguson}, {Ibata}, {Lewis}, {Richer}, \&
  {Tanvir}}]{m:08}
{McConnachie}, A.~W., {Huxor}, A., {Martin}, N.~F., {Irwin}, M.~J., {Chapman},
  S.~C., {Fahlman}, G., {Ferguson}, A.~M.~N., {Ibata}, R.~A., {Lewis}, G.~F.,
  {Richer}, H., \& {Tanvir}, N.~R. 2008, \apj, 688, 1009

\bibitem[{{Navarro} {et~al.}(1997){Navarro}, {Frenk}, \& {White}}]{NFW}
{Navarro}, J.~F., {Frenk}, C.~S., \& {White}, S.~D.~M. 1997, \apj, 490, 493

\bibitem[{{Ruiz de Austri} {et~al.}(2006){Ruiz de Austri}, {Trotta}, \&
  {Roszkowski}}]{Austri06}
{Ruiz de Austri}, R., {Trotta}, R., \& {Roszkowski}, L. 2006, Journal of High
  Energy Physics, 5, 2

\bibitem[{{Simon} \& {Geha}(2007)}]{sg07}
{Simon}, J.~D., \& {Geha}, M. 2007, \apj, 670, 313

\bibitem[{{Springel} {et~al.}(2008){Springel}, {White}, {Frenk}, {Navarro},
  {Jenkins}, {Vogelsberger}, {Wang}, {Ludlow}, \& {Helmi}}]{Springel08}
{Springel}, V., {White}, S.~D.~M., {Frenk}, C.~S., {Navarro}, J.~F., {Jenkins},
  A., {Vogelsberger}, M., {Wang}, J., {Ludlow}, A., \& {Helmi}, A. 2008, \nat,
  456, 73

\bibitem[{{Strigari} {et~al.}(2007{\natexlab{a}}){Strigari}, {Bullock}, \&
  {Kaplinghat}}]{Strig:07b}
{Strigari}, L.~E., {Bullock}, J.~S., \& {Kaplinghat}, M. 2007{\natexlab{a}},
  \apjl, 657, L1

\bibitem[{{Strigari} {et~al.}(2007{\natexlab{b}}){Strigari}, {Bulloc}k,
  {Kaplinghat}, {Diemand}, {Kuhlen}, \& {Madau}}]{stri:07a}
{Strigari}, L.~E., {Bulloc}k, J.~S., {Kaplinghat}, M., {Diemand}, J., {Kuhlen},
  M., \& {Madau}, P. 2007{\natexlab{b}}, \apj, 669, 676

\bibitem[{{Strigari} {et~al.}(2008{\natexlab{a}}){Strigari}, {Bullock},
  {Kaplinghat}, {Simon}, {Geha}, {Willman}, \& {Walker}}]{strig:08a}
{Strigari}, L.~E., {Bullock}, J.~S., {Kaplinghat}, M., {Simon}, J.~D., {Geha},
  M., {Willman}, B., \& {Walker}, M.~G. 2008{\natexlab{a}}, \nat, 454, 1096

\bibitem[{{Strigari} {et~al.}(2008{\natexlab{b}}){Strigari}, {Koushiappas},
  {Bullock}, {Kaplinghat}, {Simon}, {Geha}, \& {Willman}}]{strig:08b}
{Strigari}, L.~E., {Koushiappas}, S.~M., {Bullock}, J.~S., {Kaplinghat}, M.,
  {Simon}, J.~D., {Geha}, M., \& {Willman}, B. 2008{\natexlab{b}}, \apj, 678,
  614

\bibitem[{{Tollerud} {et~al.}(2008){Tollerud}, {Bullock}, {Strigari}, \&
  {Willman}}]{toll:08}
{Tollerud}, E.~J., {Bullock}, J.~S., {Strigari}, L.~E., \& {Willman}, B. 2008,
  \apj, 688, 277

\bibitem[{{Walsh} {et~al.}(2009){Walsh}, {Willman}, \& {Jerjen}}]{walsh09}
{Walsh}, S.~M., {Willman}, B., \& {Jerjen}, H. 2009, \aj, 137, 450

\bibitem[{{Willman} {et~al.}(2005){Willman}, {Dalcanton}, {Martinez-Delgado},
  {West}, {Blanton}, {Hogg}, {Barentine}, {Brewington}, {Harvanek}, {Kleinman},
  {Krzesinski}, {Long}, {Neilsen}, {Nitta}, \& {Snedden}}]{willman:05}
{Willman}, B., {Dalcanton}, J.~J., {Martinez-Delgado}, D., {West}, A.~A.,
  {Blanton}, M.~R., {Hogg}, D.~W., {Barentine}, J.~C., {Brewington}, H.~J.,
  {Harvanek}, M., {Kleinman}, S.~J., {Krzesinski}, J., {Long}, D., {Neilsen},
  Jr., E.~H., {Nitta}, A., \& {Snedden}, S.~A. 2005, \apjl, 626, L85

\bibitem[{{Wood} {et~al.}(2008)}]{Wood2008}
{Wood}, M., {et~al.} 2008, \apj, 678, 594

\end{thebibliography}

\end{document}